\documentstyle[aps,epsf,preprint]{revtex}

\def\[{\left[}
\def\]{\right]}
\def\nn{\nonumber}
\def\({\left(}
\def\){\right)}
\def\labels#1{\hbox{\hspace{.2in}$_{#1}$} \label{#1}}
\def\eq#1{(\ref{#1})}
\def\d{\delta}

\def\l{\lambda}

\def\.{\cdot}

\def\.{\!\cdot\!}
\def\bi{\begin{itemize}}

\def\ei{\end{itemize}}
\def\be{\begin{eqnarray}}
\def\ee{\end{eqnarray}}
\def\bn{\begin{enumerate}}
\def\en{\end{enumerate}}
\def\h{{1\over 2}}

\def\nn{\nonumber}
\def\l{\lambda}

\def\r2{\sqrt{2}}

\def\x{\times}
\def\labels#1{\label{#1}}

\def\eq#1{(\ref{#1})}
\def\A{{\cal A}}

\def\t{\tau}

\def\d{\ \delta}

\def\rr2{{1\over\sqrt{2}}}

\def\b#1{\overline{#1}}

\begin{document}

\title{Probing Extra Dimensions with Neutrino Oscillations}
\author{C.S. Lam}
\address{Department of Physics, McGill University\\
3600 University St., Montreal, Q.C., Canada H3A 2T8\\
Email: Lam@physics.mcgill.ca}
\maketitle

\begin{abstract}
In the braneworld scenario, gravity and neutrino oscillation can both be used
to detect the presence of an extra dimension. We argue that neutrino 
oscillation is particularly suitable if the size of the extra
dimension is small, in which case
 the signature for the extra dimension is the disappearance
of active neutrino fluxes into the bulk, caused by the destructive 
interference from the Kaluza-Klein states.
 We discuss a class of models
to illustrate this general feature.
\end{abstract}
\section{Introduction}
This article is dedicated to Prof.~Hiroshi Ezawa on the occasion of his
seventieth birthday. I met Hiroshi in the early 1960's, at the University
of Maryland. Right away it was clear that I could learn much from
him, both in physics and in mathematics. What I did not realize until later
was his administrative talent and his superb quality in leadership,
both amply demonstrated in his illustrious career. 
I would like to take this opportunity to wish Hiroshi a happy
birthday, a happy retirement, and many many happy returns.

The possible existence of extra (spatial) dimensions beyond our three was first
suggested by Theodor Kaluza in 1919, later modified by Oskar Klein in 1926.
When superstring came along, consistency requires it to live in
six extra dimensions. Unfortunately, there is  
no experimental evidence to date for the
presence of these extra dimensions.
 That may be due to the smallness of the extra dimensions, too small
even for the largest accelerators to see.
The lack of Kaluza-Klein excited states up to about 1 TeV
places an upper bound on the size
of the extra dimensions to be about $10^{-19}$ m.

Inspired by the discovery of higher-dimensional
D-branes in non-perturbative string theories \cite{POL},
where open strings are trapped,
a {\it braneworld scenario} was proposed \cite{BRANEWORLD} in which 
the Standard-Model (SM)
particles are confined to our three dimensional world, called a
3-brane. Only SM singlets such as gravitons and right-handed
neutrinos may leave our world
 to roam  in the extra-dimensional bulk. In that
scenario, SM particles have no Kaluza-Klein (KK) excited states 
simply because they cannot get into the bulk, so the upper bound of
$10^{-19}$ m placed on the size of the extra dimensions is no longer valid.
In order to probe the presence and the property of the extra dimensions, 
we must use either gravity, or the right-handed neutrinos $\nu_R$.

A deviation from the inverse-square law of gravity
can be used to detect the presence of an extra dimension. In a $3+n$ 
dimensional world, the surface area of a sphere 
of radius $r$ is proportional to $r^{2+n}$.
Hence the gravitational force  between two masses 
$m_1$ and $m_2$ is given by $G_nm_1m_2/r^{2+n}$, where $G_n$ is the Newtonian
gravitational constant in $n$ extra dimensions.
This is the case 
 when $r$ is less than the size $R$ of the extra dimensions.
Otherwise, the sphere is squashed in the
extra dimensions to a size $R$, so the surface area of the
squashed object is now proportional to $r^2R^n$. The 
resulting gravitational force
$G_n(\kappa/R^n)/r^2$ once again obeys the inverse-square law, 
where
$\kappa$ is a computable geometrical factor. If a deviation
from the inverse-square law is detected experimentally at $r<R$, then
$R$ marks the size of the extra dimension.  
Present experiments found no deviation
down to about 0.2 mm \cite{WASH}, so $R$ must be smaller than that.
However, it could be as large as 0.1 mm. If so, and if there are at least two
extra dimensions, then $G_n=G_0R_n/\kappa$ is large enough for strong
gravitational effects to be seen at TeV energies. This 
possibility led to a lot of 
excitement and many papers.

What if the size of the extra dimension is much smaller than 0.1 mm,
or, there is only one extra dimension with such a large size? In that case
gravity remains weak, and it is powerless to yield 
any information in the near future on extra dimensions. 

What
about the other probe, the right-handed neutrinos $\nu_R$?
They are assumed to be absent in the SM. In any case,
they are SM singlets, hence  {\it sterile},
 in the sense that they experience
 none of the SM
forces. How can they be detected even when they are present? 

The answer is
`mass'. If $\nu_R$'s exist, 
they would probably produce neutrino masses through the Dirac-mass coupling
$\overline{\nu_R}\nu_L$ with the left-handed neutrinos $\nu_L$. 
A good indication of the presence of a right-handed neutrino is therefore
the presence of a neutrino mass.

No mass has been detected in the tritium $\beta$-decay experiments.
This places an upper bound of 2.2 eV on the mass\cite{MAINZ} of the
electron-antineutrino, $\b\nu_e$. 
However, the pioneering neutrino
experiments led by Davis and by Koshiba, the beautiful data coming from
Super-Kamiokande, SNO, and KamLAND \cite{EXPT}, support an oscillation
explanation for the missing neutrinos
they detected. It demands at least two of three neutrinos
 to have non-zero masses. Specifically, if $M_1,M_2,
M_3$ are the masses of the three mass eigenstates, then $
\Delta M_\odot\equiv M_2^2-M_1^2\simeq
(7.5\x10^{-3}\ {\rm eV})^2$, and 
$\Delta M_{atm}^2\equiv 
|M_3^2-M_2^2|\simeq (50\x10^{-3}\ {\rm eV})^2$. There are also astrophysical
evidence suggesting the neutrino masses to be bounded above by 0.23 eV, if
they are degenerate \cite{WMAP}.

Unlike the quarks, oscillation experiments discover that
neutrinos mix strongly among themselves. Three rotation angles 
and one phase angle are needed
to describe the mixing of three left-handed fermions. In the case of quarks,
all three rotation angles are small. In the case of neutrinos, two of them
($\theta_{12}$ and $\theta_{23}$) are large and one of them ($\theta_{13}$)
is small.

Now that we know the neutrinos have a mass, we shall assume the right-handed
neutrinos to exist. Where do we find them?
Neutrino mixing
is large but quark mixing is small; neutrino masses are small but quark
masses are at least a million times larger. These differences suggest
that neutrinos are quite different from the quarks. Since the left-handed
quarks and the left-handed neutrinos behave in much the same way under the SM, that indicates the right-handed quarks and the right-handed neutrinos are
very different. Right-handed quarks are found at the SM energies,
this difference may be telling us that the right-handed neutrinos 
should be found elsewhere.

Where? The popular scenario is to assume the right-handed
neutrinos to live at very high energies. Through the seesaw
mechanism \cite{SEASAW}, this scenario explains the smallness of the neutrino
mass, though extra assumptions are needed to explain the large mixing of
neutrinos this way.

With the braneworld
scenario, there is another possibility. Quarks are confined to the
3-brane we live in, but $\nu_R$'s are free to roam in the bulk. 
That distinction might give rise to
 the qualitative difference between quarks and  neutrinos. In the rest
of this article, we shall examine this possibility more closely.

Each neutrino  in the bulk yields
 an infinite tower of KK
neutrinos in four dimensions, which we shall refer to as the
{\it bulk neutrinos}, or the bulk states.
 These neutrinos are non-chiral, containing both the
left-handed and the right-handed components. Since they originate
from the bulk, they are sterile.

To proceed further,
let us keep two general questions in mind. First, 
can the braneworld scenario explain the
difference between quarks and neutrinos in a natural way? 
That is important because that is the {\it raison d'\^ etre} for going into
extra dimensions.
Second,
do the data on neutrino oscillations even allow the extra-dimensions to exist?
The second question is relevant because solar and atmospheric neutrino data 
demand the mixing with sterile neutrinos to be small, but in the
braneworld scenario
all the bulk neutrinos are necessarily sterile.

The answer to these questions depends to some extent on the size 
$R$ of the extra dimension. In 
most of the
recent literature \cite{WEAK}, the size is assumed to be large. A size of 
$R=0.1$ mm corresponds to a characteristic energy  of $2\x10^{-3}$ eV,
putting it in the right ball park of the neutrino masses.
Using perturbation theory, 
a {\it weak} coupling between the left-handed brane (the SM) neutrinos and the
right-handed bulk neutrinos can be shown to 
produce a small neutrino mass as well as 
a small mixing with the sterile neutrinos. It however does not
explain the large neutrino mixing in a natural way, though that can be
arranged.

What if the size of the extra dimensions is much smaller than 0.1 mm? Then the
weak-coupling assumption explains nothing, 
so neutrino oscillation is no longer
a useful probe for extra dimensions. Neither is gravity 
in that case.
We are then back to the unenviable position
of having no way to tell the presence or absence of an extra dimension.

That however is based on the assumption that
 the brane-bulk coupling is weak. 
We shall argue that it is more natural to expect that coupling
to be {\it strong}, not weak. In that case, things are quite different,
and neutrino oscillation can provide useful information
 even when the extra dimensions are small.

Strong coupling with the bulk may
induce a large mixing between the brane neutrinos, thereby explaining naturally
why neutrino mixing can be large while quark mixing is small. 
This answers the first question posted before in the affirmative.
But why then is one of the three neutrino mixing angles
small, instead of being all large? 
We shall show that the smallness of that particular mixing angle is
intimately related to the smallness of the mass-gap ratio  
$\Delta M_\odot^2/\Delta M_{atm}^2$.

What about the second question? With a strong coupling surely we expect
a large amount of sterile neutrinos showing up in the 
the solar and atmospheric data, contrary to observation. 
Is there any way out of this fatal problem?

There is. The details will be discussed in the rest of this article,
but let us summarize here what is involved.

The trick is to introduce an extra sterile neutrino on the brane,
so that most of the mixings are between the sterile brane and bulk
neutrinos, and not between the active and sterile neutrinos.
In the strong coupling limit, the introduction of this sterile
brane neutrino is {\it forced on us} by the dynamics; it is not
arbitrary and artificial. This is so because
one of the brane neutrinos is always absorbed 
by the KK tower of bulk neutrinos in the strong coupling limit. 
In other words, if we start with
$f$ flavor neutrinos in the brane, in the strong coupling limit there are
only $f-1$ mass eigenstates left on the brane.

Therefore, in order to have two separate mass gaps needed to explain the
solar and atmospheric neutrino experiments, we need to have
three mass eigenstates in the brane,
and hence four favor states to start out with.
We know from the $Z^0$ width that there are only three active neutrinos,
$\nu_e,\nu_\mu$, and $\nu_\tau$, so the fourth one
must be sterile. We shall denote it by $\nu_s$.

With the extra dimensions small, the small mass of these neutrinos 
can no longer be explained in the usual way \cite{WEAK},
so seesaw or some other mechanism must be invoked. 
In what follows we shall not ask the origin of these small
masses, we simply introduce four parameters $m_a$  to describe the
Majorana masses of the flavor neutrinos on the brane. 

All masses from now on are understood to be
measured in some unit $U$, so the parameters $m_a$
and later on $d_a$ are dimensionless.

In addition to the four brane neutrinos, 
a minimal model would consist of a single massless bulk neutrino
in five spacetime dimensions. It decomposes into an infinite tower
of four-dimensional KK neutrinos, to be called the {\it bulk neutrinos}, or
bulk states. With $U$ properly chosen,
the spectrum of this infinite tower can be taken to be the set of all integers.
  
A Dirac mass coupling of the type
$d\overline{\nu_R}\nu_L+h.c.$ is  introduced to couple the left-handed
brane neutrinos $\nu_L$ to the right-handed bulk neutrinos $\nu_R$.
Since the KK tower comes from a single neutrino in the bulk, there 
is only one coupling constant $d_a$ per brane neutrino.

If the Dirac masses $d_a$ are comparable to 
the Dirac mass of any of the charged fermions, 
and if the Majorana masses $m_a$ are comparable to the
neutrino masses, then $d_a$
is larger than $m_b$ by more than a million times. Defining the overall
coupling strength to be $d^2=\sum_{a=1}^4d_a^2$, and letting $e_a=d_a/d$,
it is therefore likely that Nature is operating in the strong-coupling regime,
where $d\gg m_a,e_a,1$.

In summary, there are eight real parameters in the minimal theory. 
Four $m_a$'s,
and four $d_a$'s. 
The flavor neutrinos are the states when $d=0$; 
the mass eigenstates in the strong-coupling limit are
the neutrinos when $d\to\infty$. 

To get the mass eigenstates for $d\not=0$, we have to diagonalize an
infinite dimensional matrix, whose rows and columns are labeled by
the four flavor neutrinos on the brane, and the infinite number of
flavor neutrinos of the bulk. The neat thing is that 
 the eigenvalues and the eigenvectors of this infinite
dimensional matrix take on a very simple form in the strong coupling limit.

As mentioned before, one brane neutrino is absorbed into the KK tower of 
bulk neutrinos. Taking this into account, the final mass spectrum 
in the strong coupling limit
is as follows. The whole bulk spectrum is rigidly shifted by half a unit, 
so that
the masses are now half integers. The three mass eigenvalues $M_1,M_2,M_3$ on
the brane are sandwiched between the four Majorana masses $m_a$ of the
flavor neutrinos. 
Namely, if we order the parameters according to 
$m_1<m_2<m_3<m_4$, and $M_1<M_2<M_3$, then $m_1<M_1<m_2<M_2<m_3<M_3<m_4$.
We shall refer to this inequality as the {\it ordering relation}. It is a
crucial feature of the strong coupling model.

 In the strong coupling limit, we are left
with seven free parameters, four $m_a$'s and three independent $e_a$'s
(because $\sum_{a=1}^4e_a^2=1$). It turns out that
we can replace them
 by the four
$m_a$'s and the three $M_i$'s, 
{\it provided} the ordering relation is maintained. Technically
this is quite important because it is much simpler to
deal with the latter set of parameters than the former set.

The eigenvalues, and hence the unitary overlapping matrix between the flavor
and the mass eigenstates, can also be worked out.

One can then compute
the probability amplitude of a flavor neutrino $\nu_a$ oscillating into
a flavor neutrino $\nu_b$, after traversing a distance $L$. To do so,
we must 
decompose the flavor neutrino $\nu_a$ into a linear combination of the
eigenstates, because it is these normal states that propagate with a 
definite frequency. After a distance $L$,
the mass eigenstates must all be converted back into the flavor state $\nu_b$
to get the probability amplitude.

When $d$ is large, the flavor states on the brane have only a tiny 
overlap with
each of the bulk eigenstates. Nevertheless, since there are an infinite 
number of bulk eigenstates that a strongly coupled flavor brane neutrino
can reach into, the total effect of the bulk is not negligible. 
The contribution from the infinite number of
bulk states destructively interfere with one
another, so completely 
that any active neutrino that oscillates into the bulk will
not be able to come back. In other words, the bulk acts like an absorber
to the active neutrinos in the brane.

Oscillations that go through the three brane eigenstates act just like ordinary
oscillations without the presence of extra dimensions.

In other words, the effect of the extra dimensions is to cause part of the
oscillating flux of active neutrinos to be lost in the bulk.
This then is the signature of the presence of an extra dimension
 that we should look for.
 
We can now describe the physical
significance of the seven parameters in the theory. $M_1,
M_2,M_3$ are the eigenmasses of the three active brane neutrinos.
Since the mass is actually $M_iU$, it is only these products
that  can be determined experimentally. Different values  of $U$
corresponds to different size $R$ of the extra dimension, because $U$
was chosen to make the flavor mass of the bulk neutrinos 
to be an integer. Since $U$
alone cannot be determined from the experiment, neither can $R$.

This is not to say that we can use this method to probe an extra dimension
no matter how small it is. 
The strong-coupling requirement $d\gg1$ places a limit
how small $R\sim U^{-1}$ can be, if we assume 
the Dirac mass $dU\sim d/R$ to be comparable
to the Dirac mass of the charged 
fermions. The precise value of course depends
on what we use for $d$. Let us illustrate it with two extremes.
If $d/R$ is the electron mass 0.5 MeV, then we need to have $R\gg 4\x 10^{-13}$
m. If it is the top quark mass 175 GeV, then we can go down to
an $R\gg 10^{-18}$ m.

The other four parameters, $m_1,m_2,m_3,m_4$ can be used to fit the three
mixing angles, and the amount of absorption by the bulk. 
Note that there is only one free parameter to describe the potential
absorption for any $\nu_a$ oscillating into any $\nu_b$, 
so there are predictions that can be potentially falsified.

No absorption has been detected in the present data. Refined and precise
data in the future may. It that happens, it is a good indication that
an extra dimension exists.

In the minimal medel, the absence of absorption  can be achieved by letting
$m_4\to\infty$. In that limit the mixing between the active
neutrinos $\nu_e,\nu_\mu,\nu_\tau$ and the sterile neutrino
$\nu_s$ also disappears. Then we revert  to a situation 
indistinguishable from the case without extra dimensions. 
In principle, there is actually a way that might tell them apart, because the
three parameters $m_1,m_2,m_3$ in the minimal model are constrained
by the neutrino masses $M_1$ and $M_2$ through the ordering relation.
As such  we may not be able to use them to fit the three 
experimentally measured mixing angles. In reality these constraints are
fulfilled in the fit, so these two cases become indistinguishable.

When $M_1$ approaches $M_2$, the ordering constraint requires $M_1=m_2=M_2$.
This pinching of $m_2$ implies $\theta_{13}=0$. So in this model,
one obtains the interesting prediction that $\Delta M_\odot^2=0$
implies $\theta_{13}=0$. The smallness of $\theta_{13}$ is then related to the
smallness of the mass gap ratio $\Delta M_\odot^2/\Delta M_{atm}^2$.

This minimal model in five spacetime dimensions can be 
substantially generalized
without changing any of the crucial features discussed above.

These descriptions and conclusions will be put into mathematical formulas
in the next few sections. In Sec.~2, the mass matrix, its eigenvalue, and
its eigenvectors of the minimal model
are examined. They are then used to compute the 
oscillating amplitude in Sec.~3. Generalization beyond the minimal model
will be discussed in Sec.~4. 

\section{The Minimal Model and Its Solution}
The mathematical solution of the minimal model  will
be sketched here. For more details, please consult Refs.~\cite{LAM1} and \cite{LAM2}.

This model contains four flavor neutrinos in the brane, and a 
single massless flavor neutrino in a five dimensional spacetime. The latter
decomposes into a KK tower of 
bulk neutrinos with integer masses, and the former are each given a
Majorana mass $m_a\ (a=1,2,3,4)$. The brane neutrinos are coupled to the
bulk neutrinos by a Dirac-mass coupling, with coupling strengths $d_a$.
We assume all masses to be expressed in some common unit $U$, so that
the eight real parameters $m_a$ and $d_a$ are dimensionless. Direct coupling
between the brane neutrinos is assumed to be absent, and 
CP violation
is ignored in this simple model.

The symmetric mass matrix of this model is
\be
{\cal M}=\pmatrix{m&D\cr D^T&B\cr},\labels{mm}\ee
where $m={\rm diag}(m_1,m_2,m_3,m_4)$ is the $4\x 4$ mass 
matrix of the brane neutrinos, and $B={\rm diag}(0,+1,-1,+2,-2,+3,-3,\cdots)$
is the infinite dimensional mass matrix of the bulk neutrinos. 
The coupling between the two is supplied by $D$, a $4\x\infty$ matrix
in which every element of the $i$th row is equal to $d_a$.

The eigenvalue equation for the mass matrix is
\be
{\cal M}\pmatrix{w\cr v\cr}=\l\pmatrix{w\cr v\cr},\labels{meigen}\ee
where $w$ is a $4$-dimensional column vector with components $w_a$, 
and $v$ is an $\infty$-dimensional column vector with components $v_n$. $\l$
is the mass eigenvalue. In component form, \eq{meigen} reads
\be
m_aw_a+d_aA&=&\l w_a,\labels{eiv1}\\
b+(Bv)_n&=&\l v_n,\labels{eiv2}\ee
where
\be
A&=&\sum_nv_n,\nn\\
b&=&\sum_{a=1}^4d_aw_a.\labels{b}\ee
We shall choose the normalization of the eigenvectors by setting $b=1$.

The eigenvector components can be solved from \eq{eiv1} and \eq{eiv2} to be
\be
v_n&=&{1\over\l-n},\nn\\
w_a&=&A{d_a\over \l-m_a}=(Ad){e_a\over\l-m_a},\labels{eivec}\ee 
where
\be
d^2&=&\sum_{a=1}^4d_a^2,\nn\\
e_a&\equiv&d_a/d,\qquad\Rightarrow\nn\\
1&=&\sum_{a=1}^4e_a^2.\labels{e2}\ee
The constant $A$ may now be computed to be
\be
A=\sum_nv_n=\sum_m{1\over\l-m}={\pi\over\tan(\pi\l)}.\labels{a}\ee
The eigenvalue  
equation is obtained from \eq{a} and \eq{eivec}
and the normalization condition $b=1$ to be
\be
1&=&\sum_{a=1}^4d_aw_a=Ad^2\sum_{a=1}^4{e_a^2\over\l-m_a},\qquad\Rightarrow
\nn\\
{1\over\pi}\tan(\pi\l)&=&d^2\sum_{a=1}^4{e_a^2\over\l-m_a}\equiv d^2r(\l).
\labels{eiv5}\ee

Let us solve this equation for the flavor eigenvalue ($d=0$), and for the
mass eigenvalue  in the strong coupling limit ($d\to\infty$). For $d=0$,
it follows from \eq{eiv5} that $\tan(\pi\l)=0$, which implies $\l\in{\bf Z}$,
unless $\l=m_a$ for some $a$. These are the expected eigenvalues
 because they are simply
the matrix elements of the diagonal matrix ${\cal M}$ when $d=0$.

For $d\to\infty$, we should have $\tan(\pi\l)=\infty$, which implies
$\l={\bf Z}+\h$, unless $r(\l)=0$. The former are the bulk eigenvalues,
and the solutions of the latter are the brane eigenvalues
$M_i\ (i=1,2,3)$. Since $r(\l)$ approaches
$\pm\infty$ when $\l\to m_a+ 0^\pm$, there is one zero of $r(\l)$
between each successive pairs of $m_a$'s. In other words, the ordering
relation $m_1<M_1<m_2<M_2<m_3<M_3<m_4$ mentioned in the Introduction 
is obeyed.

If $d$ is large but not infinite, the eigenvalues will shift somewhat, but
they are still bounded between consecutive $m_a$'s or consecutive $n$'s.

We will now show that the three independent parameters $e_a^2$
may be replaced by the three independent parameters $M_i^2$, provided
the ordering relation holds. The argument is based on the simple observation
that $r(\l)$ is a meromorphic function of $\l$, with four simple poles occuring
at $\l=m_a$, and three zeros occuring at $\l=M_i$. 
Moreover, $r(\l)$ approaches $1/\l$ when $|\l|\to\infty$. Hence we can
write $r(\l)=\prod_{i=1}^3(\l-M_a)/\prod_{a=1}^4(\l-m_a)$. The residue at
$\l=m_b$ is then 
\be
e_b^2=\prod_{i=1}^3(m_b-M_i)/\prod_{a\not=b}(m_b-m_a).\labels{eb2}
\ee
 This formula determines
$e_a^2$ once $m_a$ and $M_i$ are known. To keep $e_a^2>0$,
the ordering relation has to be obeyed.

Let $U_\l$ be the normalized eigenvector, with components
$U_{a\l}=w_a/N$ and $U_{n\l}=v_n/N$. The norm $N^2$ of the original eigenvector
$(w_a,v_n)$ is given by $N^2=(Ad)^2s+T$, where
\be
s&=&{1\over (Ad)^2}\sum_{a=1}^4w_a^2=\sum_{a=1}^4{e_a^2\over(\l-m_a)^2},\nn\\
T&=&\sum_nv_n^2={1\over(\l-n)^2}.\labels{qt}\ee

\section{Oscillation Amplitude}
Using \eq{eivec} and \eq{qt}, 
we can calculate the transition amplitude $\A_{ab}$
from a brane neutrino of flavor $b$ and energy $E$ (measured in units of $U$),
  to a brane neutrino of flavor $a$
after it has traversed a distance $L=2E\t$ (measured in units of $U^{-1}$). 
The transition amplitude  is determined by the formula
\be
\A_{ab}(\t)&=&\sum_\l U^*_{a\l}U_{b\l}e^{-i\l^2\t}\equiv \A^S_{ab}(\t)
+\A^K_{ab}(\t),\labels{ta}\ee
where $\A^S$ is the contribution from the brane eigenvalues
$\l=M_1,M_2,M_3$, and $\A^K$ is the contribution from the bulk eigenvalues $\l\in {\bf Z}+\h$.

When $d\to\infty$, the quantities 
$v_n, w_a/Ad, s$ and $T$ are all of order 1, so
the magnitude of $w_a$ is determined by $Ad$ and the magnitude of 
$N^2$ is determined by $(Ad)^2$. According to \eq{eiv5}, $Ad=1/(dr)$.
For bulk eigenvalues, $r=O(1)$, so $Ad=O(1/d)$. This implies $N^2\simeq T$
and $U_{a\l}=O(1/d)$. In that case the bulk components of an eigenvector
are much larger than the brane components. 
For brane eigenvalues, $A=O(1)$, hence $Ad=O(d)$ and $w_a=O(d)$. In that case
the brane components of an eigenvector dominate and $N^2\simeq (Ad)^2s$.

Let us denote the large-$d$ value of $U_{aM_i}$ by $V_{ai}$, 
the value of $s$ at $\l=M_i$ by $s_i$, and
$1/(M_i-m_a)$ by $x_{ai}$. Then
\be
V_{ai}={e_ax_{ai}/\sqrt{s_i}}\quad(1\le a\le 4,\ 1\le i\le 3),\labels{vai}\ee
and
\be
\A^S_{ab}(\tau)=\sum_{i=1}^{3}V^*_{ai}V_{bi}e^{-iM_i^2\,\t}.\labels{asv}\ee
As it stands, $V$ is a $4\x 3$ matrix, but we can make it into a
square $4\x 4$ matrix by letting the last column to be $V_{af}=e_a$.
The meaning of this last column will be discussed later.
Note that we can write $V_{a4}$ in the same form as the
other $V_{ai}$, namely, $V_{a4}=e_ax_{a4}/\sqrt{s_4}$, provided we
let $\l=\infty$.

The resulting $4\x 4$ matrix $V$ can be shown to be
 real orthogonal. It depends on the seven parameters,
$m_a$ and $M_i$. 

We have shown in \eq{eb2} how to express $e_b^2$ in 
terms of these
parameters. Similarly, it can be shown that
\be
s_i=-\prod_{k\not=i}^3(M_i-M_k)/\prod_{c=1}^4(M_i-m_c).\labels{qa}\ee

We turn to the contribution from the bulk eigenvalues. Since $U_{a\l}$
is unitary, it follows from \eq{ta} that $\A_{ab}(0)=\d_{ab}$, hence
\be
\A^K_{ab}(0)=\d_{ab}-\A^S_{ab}(0).\labels{ak0}\ee
Using \eq{asv} and the unitarity of the matrix $V$, we conclude
that
\be
\A^S_{ab}(0)=\d_{ab}-V_{af}^*V_{bf}=\d_{ab}-e_ae_f.\labels{as0}\ee
Therefore
\be
\A^K_{ab}(0)=e_ae_b.\labels{ak01}\ee

The contribution from the bulk eigenvalues can also be obtained directly
from \eq{ta} and the paragraph following that equation to be
\be
\A^K_{ab}(\t)&=&\sum_{\l\in {\bf Z}+\h}{1\over (dr)^2T} 
{e_ae_b\over(\l-m_a)(\l-m_b)}e^{-i\l^2\t}\equiv e_ae_bF(\t).\labels{aa}\ee
Both $r$ and $T$ are of order 1 as $d\to\infty$, so the contribution
from each bulk eigenvalue to the sum is $O(1/d^2)$. Since there
are an infinite number of bulk eigenvalues, the total contribution to the
sum in \eq{aa} is not necessarily zero. In fact, we know 
from \eq{ak01} that $F(0)=1$ even at an infinite $d$. 

It can be shown that $F(\t)=g(K^2\t)$, where
$K^2=d^2(1+\pi^2d^2)$, and that
\be
g(x)\equiv {1\over\pi}\int_{-\infty}^\infty{e^{-iu^2x}\over u^2+1}\labels{g}\ee
is zero at $x=0$, and decrease to 0 like  $(1-i)/\sqrt{2\pi x}$ for large
$x$. This means that ${\cal A}^K_{ab}(\t)=e_ae_bg(K^2\t)$
is zero whenever $\t>0$, in the limit $d\to\infty$. 
This function describes the absorption of the active neutrino flux
into the bulk.

Therefore, in the strong coupling limit, we end up with
\be
{\cal A}_{ab}(\t)={\cal A}^S_{ab}(\t)=\sum_{i=1}^{3}V^*_{ai}V_{bi}e^{-iM_i^2\,\t}.
\labels{amp4}\ee

In the limit $m_4\to\infty$, it follows from \eq{eb2} that $e_4^2\to 1$,
and hence from \eq{e2} that $e_b\to 0$ for $b=1,2,3$. Since the matrix
$V$ is real orthogonal, and $e_4=V_{44}$, it also follows that
$V_{a4}=0$ for $a=1,2,3$. As a result, the active neutrinos 
$\nu_e,\nu_\mu,\nu_\tau$ 
do not mix with the sterile neutrino $\nu_s$, and the active neutrinos
do not get absorbed by the bulk.

If furthermore $M_1=M_2$, then the ordering relation forces $m_2=M_1$.
In that case $V_{23}=0$. This means the mixing angle $\theta_{13}=0$
if we identify the $a=2$ flavor neutrino
with $\nu_e$. Hence a vanishing $\Delta M_\odot^2/\Delta M_{atm}^2$ implies
a vanishing $\theta_{13}$.

\section{Generalization of the Minimal Model}
The final result \eq{amp4}  remains valid for almost all
$B$ in \eq{mm}. This is reasonable because \eq{amp4} does not depend
on the property of the absorptive bulk, which $B$ affects. For a detailed
argument, please consult Ref.~\cite{LAM3}.

This research is supported by NSERC and FRNT.


\end{document}